\documentclass{PoS}

\usepackage[style=numeric-comp,maxnames=3,backend=bibtex8,doi=false,isbn=false,url=false,sorting=none,sortcites=true]{biblatex}
\AtEveryBibitem{\clearfield{title}}

\usepackage{amsmath}    
\usepackage{graphicx}   
\usepackage{verbatim}   
\usepackage{xspace}
\usepackage{xcolor}

\usepackage{amssymb}

\usepackage{bm}
\usepackage{esvect}
\usepackage{gensymb} 

\newcommand{\eg}{e.g.\/,\xspace}

\newcommand{\Sim}{\sim\kern-0.2em\xspace}

\newcommand{\txt}[1]{\text{#1}}

\newcommand{\python}{\texttt{Python}\xspace}

\newcommand{\ctools}{\texttt{ctools}\xspace}
\newcommand{\ctlike}{\texttt{ctlike}\xspace}
\newcommand{\gammapy}{\texttt{Gammapy}\xspace}

\newcommand{\dgr}{\ensuremath{\degree}\xspace}

\newcommand{\fermi}{{\textit{Fermi}}\xspace}
\newcommand{\fermil}{{\textit{Fermi}-LAT}\xspace}
\newcommand{\fermig}{{\textit{Fermi}-GBM}\xspace}

\newcommand{\Swift}{\textit{Swift}\xspace}
\newcommand{\swift}{\textit{Swift}\xspace}
\newcommand{\swiftb}{\textit{Swift}-BAT\xspace}

\newcommand{\cta}{{CTA}\xspace}

\newcommand{\pl}{PL\xspace}

\newcommand{\grb}{\ensuremath{\txt{GRB}}\xspace}
\newcommand{\grbs}{\ensuremath{\txt{GRBs}}\xspace}

\newcommand{\gamray}{\ensuremath{\gamma\txt{-ray}}\xspace}

\title{POSyTIVE - a GRB population study for the Cherenkov Telescope Array}

\ShortTitle{POSyTIVE - a GRB population study for CTA}

\author{
Maria Grazia Bernardini,$^{1}$
Elisabetta Bissaldi,$^2$
Zeljka Bosnjak,$^3$
Alessandro Carosi,$^4$
Paolo D'Avanzo,$^1$
Tristano Di Girolamo,$^5$
Susumu Inoue,$^6$
Thomas Gasparetto,$^{4,7}$
Giancarlo Ghirlanda,$^1$
Francesco Longo,$^7$
Andrea Melandri,$^1$
Lara Nava,$^{1,7}$
Paul O'Brien,$^8$
\speaker{Iftach Sadeh},$^{9}$
Fabian Sch\"ussler,$^{10}$
Thierry Stolarczyk,$^{11}$
Susanna Vergani,$^{12}$
Carlo Francesco Vigorito,$^{13}$
on behalf of the CTA Consortium\footnote{~For the consortium list, see PoS(ICRC2019)1177~;~~See also: \protect\url{https://www.cta-observatory.org/}~.}
\\
\noindent{$^1$}INAF-Osservatorio Astronomico di Brera, Via Bianchi 46, I-23807 Merate (LC), Italy;\hspace{10pt}
{$^2$}Politecnico and INFN, Bari, Italy;\hspace{10pt}
{$^3$}Faculty of Electrical Engineering and Computing, University of Zagreb, Zagreb, Croatia;\hspace{10pt}
{$^4$}Laboratoire d'Annecy de Physique des Particules, Univ. Grenoble Alpes, Univ. Savoie Mont Blanc, CNRS, LAPP, 74000 Annecy, France;\hspace{10pt}
{$^5$}University ``Federico II'' and INFN, Napoli, Italy;\hspace{10pt}
{$^6$}RIKEN, Wako,  Japan;\hspace{10pt}
{$^7$}University of Trieste and INFN-Trieste, Via Alfonso Valerio, 2, 34127, Trieste, Italy;\hspace{10pt}
{$^8$}School of Physics and Astronomy, University of Leicester, University Road, Leicester, LE1 7RH, United Kingdom;\hspace{10pt}
{$^{9}$}DESY, Platanenallee 6, 15738 Zeuthen, Germany;\hspace{10pt}
{$^{10}$}IRFU, CEA, Universit\'e Paris-Saclay, F-91191 Gif-sur-Yvette, France;\hspace{10pt}
{$^{11}$}AIM, CEA, CNRS, Universite Paris-Saclay, Universite Paris Diderot, Sorbonne Paris Cite, F-91191 Gif-sur-Yvette, France;\hspace{10pt}
{$^{12}$}GEPI, Observatoire de Paris, PSL University, CNRS, Meudon, France;\hspace{10pt}
{$^{13}$}University and INFN Torino, via Pietro Giuria 1, 10125 Torino, Italy.\\
E-mail: \email{iftach.sadeh@desy.de}
}

\abstract{
One of the central scientific goals of the next-generation Cherenkov Telescope Array (CTA) is the detection and characterization of gamma-ray bursts (GRBs). CTA will be sensitive to gamma rays with energies from about 20\,GeV, up to a few hundred TeV. The energy range below 1\, TeV is particularly important for GRBs. CTA will allow exploration of this regime with a ground-based gamma-ray facility with unprecedented sensitivity. As such, it will be able to probe radiation and particle acceleration mechanisms at work in GRBs. In this contribution, we describe POSyTIVE, the POpulation Synthesis Theory Integrated project for very high-energy emission. The purpose of the project is to make realistic predictions for the detection rates of GRBs with CTA, to enable studies of individual simulated GRBs, and to perform preparatory studies for time-resolved spectral analyses.
The mock GRB population used by POSyTIVE is calibrated using the entire 40-year dataset of multi-wavelength GRB observations. As part of this project we explore theoretical models for prompt and afterglow emission of long and short GRBs, and predict the expected radiative output. Subsequent analyses are performed in order to simulate the observations with CTA, using the publicly available \ctools and \gammapy frameworks. We present preliminary results of the design and implementation of this project.
}

\FullConference{36th International Cosmic Ray Conference -ICRC2019-\\
		July 24th - August 1st, 2019\\
		Madison, WI, U.S.A.}

\begin{document}

\section{The POSyTIVE project}

Gamma-ray bursts (\gamray; \grbs) are among the most powerful phenomena in the Universe.
They are appealing targets for
the upcoming Cherenkov Telescope Array (\cta), 
especially considering the recent
detections at very high-energies~\cite{Acharya:2017ttl, 2019ATel12390....1M, hessGrbAfterglow}.
Observations with \cta will provide valuable insights into particle acceleration at relativistic shocks and radiative processes at work in GRBs. They will enable the use of GRBs as tools to investigate fundamental physics questions (such as violations of Lorentz invariance), and to probe the extragalactic background light.

To perform preparatory studies in view of CTA and predict the CTA GRB detection rate, we are developing the POSyTIVE\footnote{~POSyTIVE, POpulation SYnthesis Theory Integrated model for Very high Emission.} project. 
The  project will combine population models for both short and long GRBs with emission models of prompt and afterglow radiation, and simulate the detectability of the predicted emission with the CTA.

This method will provide a library of simulated GRBs, which can then be used to i) test follow up strategies of CTA, taking account of the sensitivity on the appropriate time scales; and
ii) study which region of the physical parameter space of the GRB population can be constrained by future CTA observations. 

\begin{figure*}
    \centering
    \includegraphics[scale=0.5]{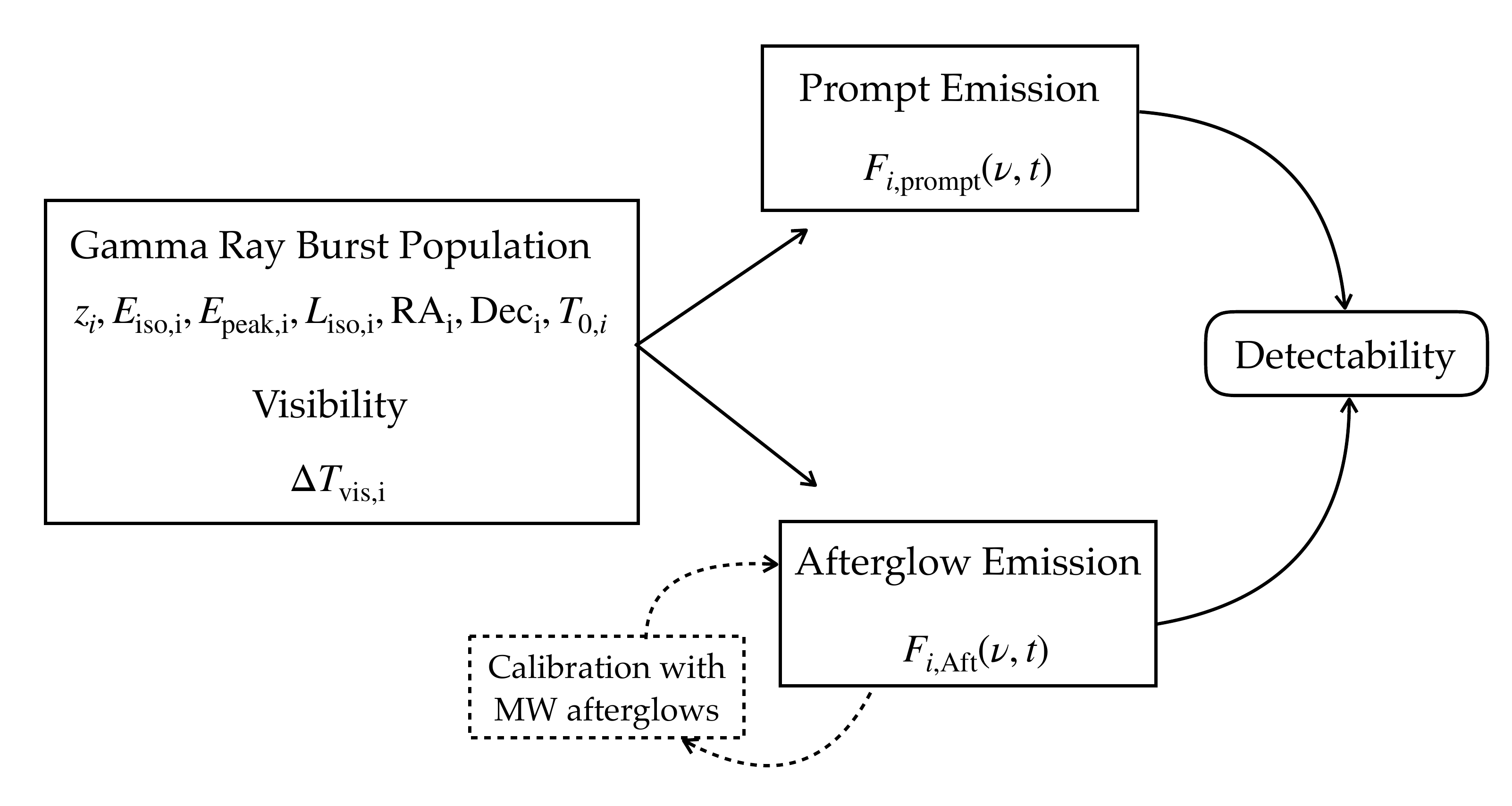}
    \caption{Schematic representation of POSyTIVE. Boxes show the modules implemented in the project, highlighting the outcome of each element. The intrinsic \textit{population} properties from the population synthesis are input parameters for the afterglow \textit{emission} modules.
    These parameters are further used to select, among the already simulated set of prompt emission models, those that are consistent with 
    synthetic populations. 
    The \textit{visibility} represents the time window when each simulated burst is observable from the two CTA sites, accounting for additional observational constraints. }
    \label{fig1}
\end{figure*}

\section{The population of short and long GRBs}\label{sec:population}

The synthetic GRB populations are simulated based on a minimal set of intrinsic properties, following \cite{Ghirlanda2016,Ghirlanda2015}. Here, we describe the method for the population of long GRBs. A similar scheme is adopted for the population of short bursts.  We make the following assumptions: 
\begin{itemize}
    \item broken power law distribution of the rest frame peak energy, $E_{\rm peak}$, of GRBs. 
    The Monte Carlo (MC) sampling of this distribution provides for each simulated GRB an intrinsic $E_{\rm peak,i}$;
    \item GRB redshift distribution prescribing the event rate density (in units of Gpc$^{-3}$ yr$^{-1}$). The long GRB formation rate is assumed to be proportional to the cosmic star formation rate~\cite{Madau2014}, with the additional possibility of its evolution with redshift.
    This provides $z_{\rm i}$;
    \item GRBs follow the empirical correlations between $E_{\rm peak}$ and the isotropic equivalent energy, $E_{\rm iso}$, and luminosity, $L_{\rm iso}$~\cite{Amati2002, Yonetoku2004}.  
    These provides $E_{\rm iso,i}$ and $L_{\rm iso,i}$.
\end{itemize}

In addition, we obtain the distribution of observer frame afterglow onset times from the available measurements and lower limits~\cite{Ghirlanda2018}.  This observable allows us to infer $\Gamma$, the bulk Lorentz factor of GRBs during the coasting phase. From the onset time, we obtain $\Gamma_{0,{\rm i}}$ for either the case of a constant density circumburst medium, or for a wind density profile.

In order to calibrate the simulated populations with the observed GRB samples, we assume an intrinsic spectrum. The latter is used to compute the flux and fluence (respectively corresponding to $L_{\rm iso,i}$ and $E_{\rm iso,i}$) in the  energy bands corresponding to the instruments, which provide the real GRB sample as constraints. We assume that all GRB spectra are described by the Band function~\cite{1993ApJ...413..281B}. The peak energy is provided as described above. The low and high energy spectral indices are derived by randomly sampling Gaussian distributions.

For each GRB, whose intrinsic properties are drawn via MC from the above probability density distribution, we derive the observer frame properties. The latter include, \eg the flux and fluence in a given energy band, $\Delta E$. The free parameters of the simulated population are then constrained by observational data. 
We explore the parameter space in search of combinations that produce a simulated GRB population, consistent with the following: 
\begin{enumerate}
    \item \textbf{The GRB population detected by \fermig.} We further limit this comparison to the bright end of the \fermi population in order to avoid possible selection bases affecting the faint end of the peak flux distribution. We compare and reproduce with the simulated population the distributions of peak flux, fluence, duration, and observer frame peak energy of \fermig bursts. As an independent check of the simulated population, we also verify the consistency with the peak flux distribution of \swiftb detected GRBs.
    \item \textbf{The complete sample of \swift GRBs} (selection criteria described in~\cite{Salvaterra2012, Pescalli2016}). These are bright \swift GRBs (with a 15--150 keV peak flux $> 2.6$ ph cm$^{-2}$ s$^{-1}$), which have measured redshifts and prompt emission properties ($E_{\rm iso}$, $L_{\rm iso}$, $E_{\rm peak}$). We require that the simulated GRB population (with the same peak flux cut as above) matches the distributions of  $z$, $E_{\rm iso}$, and $L_{\rm iso}$ of the \Swift complete sample. The same \swift sample is also adopted (see \S4) for the calibration of the afterglow parameters. 
\end{enumerate}


The visibility of each GRB from each CTA site is evaluated using a \python code based on Astroplan~\cite{morris18}\footnote{~\url{https://astroplan.readthedocs.io/en/latest/}~.}. This is an open source \python package that, among other features, determines observability of sets of targets from a user-defined observatory, given an arbitrary set of constraints (i.e., altitude, airmass, moon separation \& illumination, etc.). Specifically, for each GRB of the population, the code checks the observability from each CTA site during the 24~hours following the trigger time. It then provides a calculation of the corresponding temporal window of visibility. The constraints to determine the observability that we adopted are the following: i) we require that the Sun is 18\dgr below the horizon (astronomical twilight); ii) we constrain the minimum altitude of the target, as 10\dgr above the horizon. 

\section{The GRB calibration sample}\label{sec:calibration_sample}

As described above, the simulated population of GRBs is compared to the observed properties of real samples of long and short GRBs, the so called \textit{BAT6} and \textit{SBAT4} samples~\cite{Salvaterra2012, Pescalli2016,Davanzo2014}. 

Both datasets are compiled by selecting GRBs detected by \Swift, having favorable observing conditions for redshift determination from the ground. These are events with low Galactic extinction in the direction of the burst, $A_V < 0.5$ mag, restricted to those that are bright in the 15-150 keV \swiftb energy band. 
In particular, this last criterion is equivalent to a threshold for the peak photon flux, $P$, measured in the 15--150 keV energy band by the \swiftb.

For the BAT6 sample, we select long GRBs with $P \geq 2.6$ ph s$^{-1}$ cm$^{-2}$, computed using the \swiftb light curves binned with $\delta t = 1$~s.
For the SBAT4 sample, we select short GRBs with $P \geq 3.5$ ph s$^{-1}$ cm$^{-2}$, computed using the \swiftb light curves binned with $\delta t = 64$~ms.
Such a high flux cut ensures that our samples are free from any detector-related threshold bias. With the above criteria, the two samples comprise 99 long GRBs (83\% having redshifts~\cite{Pescalli2016}) and 16 short GRBs (69\% having redshifts~\cite{Davanzo2014}).

Being free of selection effects (except flux limits), both samples provide the possibility to compare the rest-frame physical properties of GRB prompt and afterglow emission in an unbiased way~\cite{Salvaterra2012,Nava2012,Campana2012,Melandri2012,Davanzo2012,Covino2013,Melandri2014,Davanzo2014,Vergani2015,Japelj2016,Pescalli2016,Asquini2019,Palmerio2019}.

\section{The prompt model}

To investigate the properties of the prompt emission, we use the numerical code described in \cite{bosnjak09}. The simulated synthetic population of GRBs provides the following properties for each event: peak energy of the prompt spectrum, isotropic energy, $E_{\rm iso}$, redshift, Lorentz factor, and the duration of the prompt emission.
At this stage, we use a large set of simulated spectra, calculated in the comoving frame of the shocked material~\cite{bosnjak09,bosnjak14}. These are produced for a range of values of the magnetic field,  the dynamical timescale, the electron density, and the minimum electron Lorentz factor of the electron distribution. 

Each emitted photon spectrum is computed from the time-dependent evolution of relativistic electrons. We do not include the possible contribution from a population of accelerated relativistic protons.
Accelerated  electrons in the amplified magnetic field radiate synchrotron photons; these synchrotron photons may be scattered to higher energies by relativistic electrons (inverse Compton).
At low energies the photons may be absorbed (synchrotron self-absorption).  At high energies, we account for photon-photon annihilation, producing electron-positron pairs.\footnote{~The contribution of these pairs to the radiation is not considered in the present version of the code.} 

The resulting spectra in the comoving frame are transformed to the observer frame using the provided values of bulk Lorentz factor and redshift. The peak energy of the spectrum and the observed flux are then compared with the provided values of the synthetic GRB population. 
In this way, we determine the parameters of the simulated spectra in the comoving frame, which can reproduce the synthetic GRB population. The corresponding spectral components are derived in the energy band observable by CTA.

\section{The afterglow model}
To simulate afterglow radiation we consider a standard scenario of synchrotron and synchrotron-self Compton (SSC) radiation from electrons accelerated by the forward shock. 
To describe the dynamics of the blastwave, we follow the method developed by~\cite{NavaSironi13}.
Particle acceleration and magnetic field amplification at the shock are described 
by introducing the parameters $\epsilon_{\rm e}$ and $\epsilon_{\rm B}$; these respectively denote the fraction of shock-dissipated energy that is used to accelerate the electrons and to amplify the magnetic field.
Electrons are
assumed to be efficiently accelerated into a powr law (\pl) energy distribution, $dN(\gamma)/d\gamma\propto\gamma^{-p}$, where $\gamma$ is the electron Lorentz factor.

Synchrotron and SSC radiation spectra and their evolution with time are described following~\cite{granot02,sari01}. Possible modifications to both spectral components caused by Klein-Nishina (KN) effects follow the prescriptions in~\cite{nakar09}.

\begin{figure*}[tp]
    \begin{minipage}[c]{1\textwidth}
        \begin{minipage}[c]{1\textwidth}
          \begin{center}
            \begin{minipage}[c]{0.49\textwidth}\begin{center}
              \includegraphics[trim=0mm 0mm 0mm 0mm,clip,width=1\textwidth]{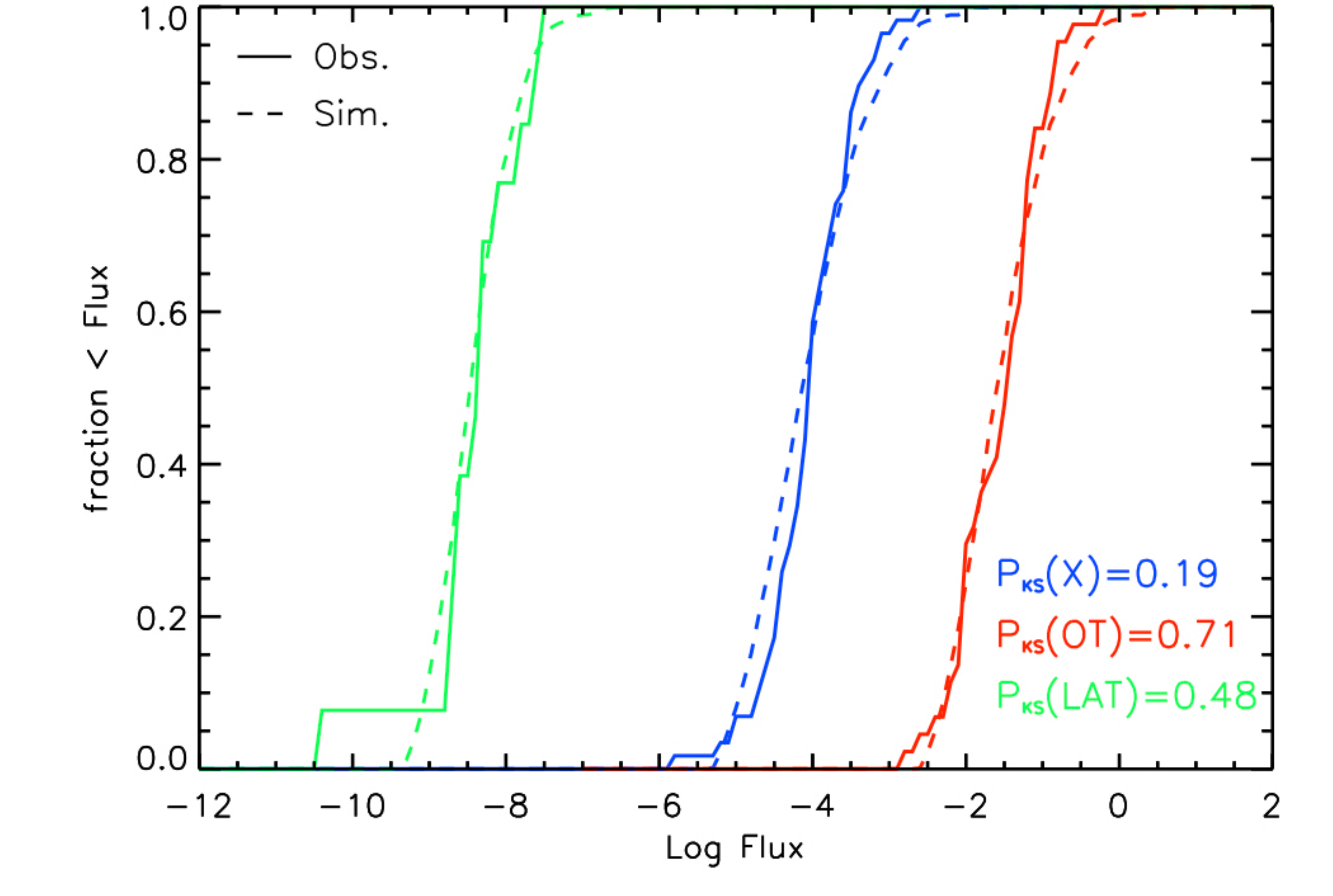}
            \end{center}\end{minipage}\hfill
            \begin{minipage}[c]{0.49\textwidth}\begin{center}
              \includegraphics[trim=0mm 0mm 0mm 0mm,clip,width=1\textwidth]{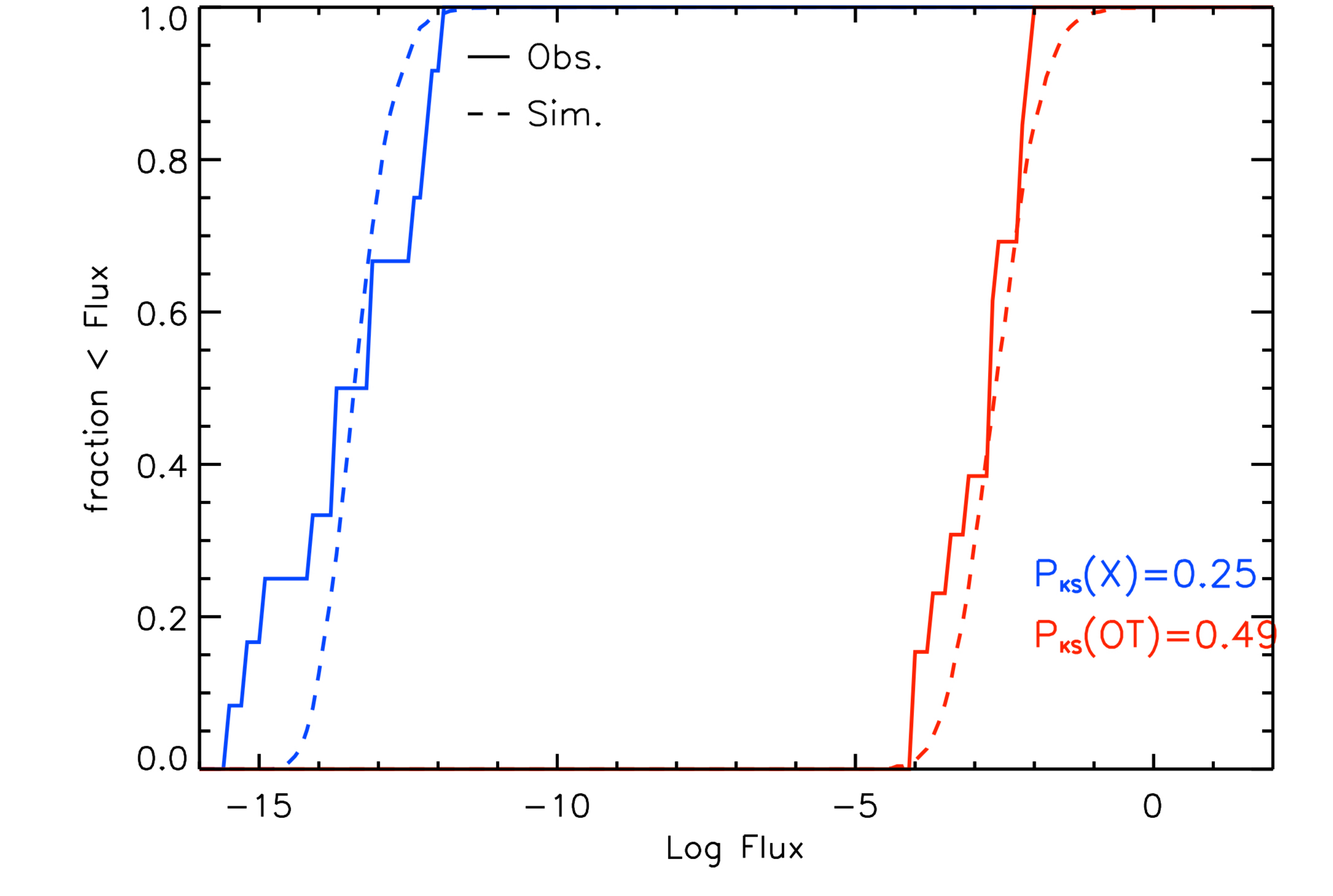}
            \end{center}\end{minipage}\hfill 
          \end{center}
        \end{minipage}\hfill
        %
      %
      \begin{minipage}[c]{1\textwidth}
        \begin{center}
          \begin{minipage}[t]{1\textwidth}\begin{center}
            \caption{\label{fig:calibration} Comparison between multi-wavelength afterglow observations of the complete sample of \Swift GRBs (solid lines), and simulations of afterglow radiation for the mock population, shown for long (left panel) and short (right panel) GRBs. Solid lines show the cumulative flux distribution for real GRBs,
            where dashed curves refer to the simulated afterglow emission.
            Here, $P_{\rm KS}$ is the probability associated to the Kolmogorov-Smirnov statistic that the two distributions are drawn from the same population.
            The curves are colour-coded: blue/red: X-ray/optical flux at 11\,hours; green: \fermil flux at 100\,s.}
          \end{center}\end{minipage}\hfill
        \end{center}
      \end{minipage}\hfill
      \vspace{15pt}
    \end{minipage}\hfill
  \end{figure*}

For each GRB of the synthetic population, the initial Lorentz factor is given by~\S\ref{sec:population}. The initial blastwave kinetic energy, $E_{\rm k}$, is inferred from $E_{\rm iso}$, assuming an efficiency, $\eta_\gamma=20\%$, for the prompt emission mechanism, $E_{\rm k}=E_{\rm iso}\,(1-\eta_\gamma)/\eta_\gamma$. 
To predict the afterglow emission for each GRB, we need to specify the values of the remaining model parameters: the density of the external medium and the corresponding radial profile, ($n(r)=n_0\,r^{-s}$), $\epsilon_{\rm e}$, $\epsilon_{\rm B}$, and $p$.
Given the different progenitors and environments of the two classes of GRBs, we assume $s=0$ for short GRBs, and $s=2$ for long ones.
The large uncertainties on the remaining free parameters are such that changing the assumed values has a strong impact and small predictive power on the resulting emission.
To overcome this problem, before predicting the emission in the CTA energy range, we calibrate the model parameters to reproduce existing afterglow observations at lower frequencies.
We use the calibration sample presented in~\S\ref{sec:calibration_sample}. We find the set of values for the free model parameters which are able to describe optical-to-GeV observations. Preliminary results of the parameter calibration are shown in Fig\/.~\S\ref{fig:calibration}.

These parameters will be used to predict the radiation in the CTA energy range for the full synthetic GRB population.
%
  %
  %

\section{Preliminary tests of detectability with \cta}
To test the detection prospects of GRBs by CTA, two  methods are being studied. The first uses a sky map likelihood maximization (as for \fermil), and has been implemented using the \ctools framework~\cite{ctools}. The second method relies on the \textit{on-off} approach, the analysis traditionally used for ground Cherenkov telescopes; it has been implemented as part of a \gammapy-based code~\cite{Gammapy}.

The tests on the simulation pipelines performed to date take as input time-variable very high-energy spectra, derived from GRB afterglows. The selected GRBs are located at different redshifts. Their spectra are properly modified by interactions with the extragalactic background light, following the model of~\cite{dominguez2013}. The GRB spectra are modelled as simple power laws for a series of fixed time bins in a logarithmic scale. Each time slice is considered as an independent observation, to which the CTA instrument response functions (IRFs)~\cite{CTAperformance}
 are applied. We consider a point-like source in the center of the field-of-view, a zenith angle of 20$\degree$, and an assumed observation time of 100 s. At each time step observations are stacked, and the detection significance is computed. The time slice in which a 3$\sigma$ or 5$\sigma$ detection with a confidence level of 90\% is obtained from $\sim$100~MC trials gives the corresponding detection time.
 
In the first analysis chain, based on the \ctools library, a sky map is produced from the \textit{full-enclosure} IRF. The latter contains the CTA response over the entire field of view ($\pm 2.5\degree$). This allows the use of a likelihood analysis (the {\ctlike} tool), assuming a constant background in time and a simple \pl spectral model. A MC simulation involves generating sky maps with varied pixel counts, assuming Poissonian statistics. The detection significance is computed from the test statistics of the likelihood fit.

The current \gammapy-based pipeline uses the CTA \textit{on-axis} point-like source IRF, for which the 68\% containment radius around a source is intrinsically optimized. For all energy bins and each time slice, the predicted signal and background are thus directly obtained (no sky map is simulated). The detection significance calculation follows~\cite{Lima1983},  with $\alpha=1/5$ (i.e. it is considered that the background is obtained from a region which is 5~times larger than the signal region). The simulation involves fluctuating the signal and background counts, based on Poissonian statistics.

Preliminary tests have been performed on a sub-sample of $\sim10$ generic GRB light-curves, with the intrinsic \grb flux scaled by various factors (between 1/2 and 1/100), testing the capability of \cta to detect faint GRBs.
A preliminary analysis of these simulations shows that the majority of the events ($>50\%$) are detected with $\geq3\sigma$ significance in both analysis chains. The likelihood analysis gives systematically higher significance values than the on-off method, since it uses the full point spread function information. However, the on-off method is much faster. It might be more suitable for situations in which the IRFs are poorly known or have large systematic uncertainties; for instance, it might be appropriate for online analyses.

The two pipelines will be used on the larger, theoretically-based, GRB population, in order to derive our final estimates for the number of potentially detectable GRBs.

\section*{Acknowledgements}
This work was conducted in the context of the CTA Transients Working Group. We gratefully acknowledge financial support from the agencies and organizations listed here: \url{http://www.cta-observatory.org/consortium\_acknowledgments}.

\printbibliography


\end{document}